\documentstyle[12pt]{article}

\begin{document}

\baselineskip=14pt plus 0.2pt minus 0.2pt
\lineskip=14pt plus 0.2pt minus 0.2pt

\begin{flushright}
Quantum Optics (to be published):  LA-UR-93-2804 \\
\end{flushright}

\begin{center}
\Large{\bf
RYDBERG WAVE PACKETS ARE SQUEEZED STATES} \\

\vspace{0.25in}

\large

Michael Martin Nieto$^a$\\
{\it
 Theoretical Division, Los Alamos National Laboratory\\
University of California\\
Los Alamos, New Mexico 87545, U.S.A. \\}

\normalsize

\vspace{0.3in}

{ABSTRACT}

\end{center}
\begin{quotation}
We point out that  Rydberg wave packets (and similar ``coherent" molecular
packets) are, in
general, squeezed states,
rather  than the
more elementary coherent states.  This observation allows a more intuitive
understanding
of their properties; e.g.,   their revivals.

\vspace{0.25in}

\noindent PACS: 03.65.-w, 02.20.+b, 42.50.-p

\end{quotation}

\vspace{0.3in}


Over the past decade, a very interesting phenomenon has been observed
experimentally
 and discussed theoretically  \cite{ryd1,ryd2,ryd3,ryd4},
Rydberg wave packets.  A short-pulsed laser beam is used to excite a
high-$\langle n
\rangle$
``Rydberg wave packet."  This packet often has a significant overlap with a
small
number of  eigenstates, it exhibits classical motion for a short period, it
disperses,
and then undergoes periodic revivals.   As such, these systems are often
referred as
being in a coherent state.  In this short letter we point out that, actually,
these
packets are composed of generalized squeezed states, as are some other coherent
quantum systems, such as laser excited molecules \cite{mole}.

To set the stage, let us quickly review some of the important properties of the
squeezed states of the harmonic oscillator \cite{bonny,ss}.  From the
wave-function
point of
view, these states  minimize the $x-p$ uncertainty relation at $t=0$. They are
Gaussians, displaced from the origin, whose width is not that of the ground
state:
\begin{equation}
\psi_{ss}(x) = \left[\frac{\omega}{\pi s^2}\right]^{1/4}
\exp\left[-\frac{\omega(x-x_0)^2}{2s^2}+ip_0x\right],   \label{Gauss}
\end{equation}
where $\hbar=m=1$.
If $s=1$,  the Gaussians have the width of the ground state of the harmonic
oscillator.
Then they  are the special case of coherent states.

Both coherent states and squeezed states follow the classical motion in that
\begin{equation}
\langle x \rangle_{cs} = \langle x \rangle_{ss} = x_{cl}
                       = x_0 \cos(\omega t) + \frac{p_0}{\omega} \sin(\omega
t).
\end{equation}
However, their uncertainty products vary with time as
\begin{equation}
[\Delta x(t)]^2 = \frac{1}{2\omega} \left[s^2 \cos^2 \omega t +
                  \frac{1}{s^2} \sin^2 \omega t \right] ,  \label{dx}
\end{equation}
\begin{equation}
[\Delta p(t)]^2 = \frac{\omega}{2} \left[\frac{1}{s^2} \cos^2 \omega t +
                  {s^2} \sin^2 \omega t \right] ,
\end{equation}
\begin{equation}
[\Delta x(t)]^2 [\Delta p(t)]^2 = \frac{1}{4} \left[1 +
                  \frac{1}{4}\left(s^2 - \frac{1}{s^2}\right)
                \sin^2[2 \omega t]\right] .
\end{equation}
Thus, although the $s=1$ coherent states have an unvarying uncertainty product
and maintain their shapes, the squeezed states do not. As Eq.
(\ref{dx}) shows, four times every  classical period a squeezed state will
return to  the
coherent-state minimum-uncertainty value for position, $1/2\omega$;
specifically when
\begin{equation}
\sin^2 \omega t = \left(\frac{s^2}{1 + s^2}\right) .
\end{equation}
For larger or smaller values of $t$, the uncertainty value will be different.
A
similar statement holds for momentum.

These are a type of ``revival," which indeed is a specialized flowering of a
general phenomenon.  It has long been known that in a system with
equally-spaced
eigenenergies, any wavepacket will reconstruct itself every classical period of
oscillation \cite{oldpac}.  This is simply because the decomposition of the
wave
function into eigenstates,
\begin{equation} \Psi(x,t) = \sum_{n=0}^{\infty} c_n
exp[-i\omega_0 t(n+n_0)] \psi_n(x) ,
\end{equation}
means the wave function will return to its original shape after every period
$t=2\pi /\omega_0 $. Gaussians, in a harmonic oscillator potential, do even
better.

 Another property of squeezed states vs. coherent
states is that they can be constructed so as to
significantly overlap fewer eigenstates (sub Poisson number distribution) or
more
eigenstates (super Poisson distribution) than a coherent state would (Poisson
distribution).  For a squeezed state (now $\omega = 1$)
\begin{equation}
\langle N \rangle = \alpha_1^2 + \alpha_2^2
        +\left[ \frac{(s^2-1)^2}{4s^2} \right] ,
\end{equation}
\begin{equation}
(\Delta N)^2 = \alpha_1^2 s^2 + \frac{\alpha_2^2}{s^2}
      +\left[\frac{(s^2+1)^2}{2s^2} \right] \left[ \frac{(s^2-1)^2}{4s^2}
\right] ,
\end{equation}
where $\alpha = \alpha_1 + i\alpha_2= x_0/\sqrt 2 + i p_0/\sqrt 2$ is the usual
coherent state parameter and $s=1$ is the limit to a coherent state.  Further,
for large squeezing, oscillations occur in the probability distribution as a
function of $n$ \cite{schl}.

Consider, in the above, the cases where $\alpha_1^2 = 0$,  $\alpha_2^2 \sim
64$, and $s^2 = 4$ or $1/4$, respectively.  Then $(\Delta N) \sim
\langle N \rangle^{1/2} /s
\sim 4$ or $16$, respectively, instead of $8$, as would be the case for a
coherent
state.   Therefore, such  states would  exhibit ``squeezing" in their evolution
and
revivals.

One can use Gaussians as approximate states to describe physical situations.
However, for more general systems, more general coherent and squeezed states
are called
for as a matter of principle.  Their advantage is to better take into account
the
different shapes of the various potentials, especially the centripetal barriers
of
angular momentum.  Because of these different shapes, packets will evolve
differently,
but one still wants to maximize the coherence properties of the wave packet.

There exists a method to obtain coherent and squeezed states for general
systems
that was inspired by Schr\"{o}dinger's original discovery of coherent states
\cite{sch1}.   Schr\"{o}dinger was interested in finding states which follow
the
classical motion and do not charge their shapes with time.  His expressed
belief
that
the same could be done for the Kepler problem led to disagreement between he
and
Heisenberg \cite{history}.  The controversy resulted in Heisenberg's discovery
of the
uncertainty relation \cite{hei} and Schr\"{o}dinger's generalization  of it
\cite{sch2}.

 In this method one starts with the classical
problem and transform it to the ``natural classical variables," $X_c$ and
$P_c$,
which vary as the $\sin$ and the $\cos$ of the classical $\omega t$ (or
$\theta(t)$, for spherical systems).   The Hamiltonian
then is  of the form $P_c^2 + X_c^2 $.  Now take these natural classical
variables and transform them into ``natural quantum operators."  Since these
are
quantum operators, they have a commutation relation and uncertainty relation:
\begin{equation}
[X,P] = iG, \hspace{0.5in}
(\Delta X)^2(\Delta P)^2 \geq {\frac{1}{4}}\langle G\rangle ^2.
\label{uncert}
\end{equation}
The states
that minimize the uncertainty relation above are  the solutions to
\begin{equation}
Y\psi_{ss} \equiv
\left(X + \frac{i\langle G\rangle }{2(\Delta P)^2} P\right)\psi_{ss}
=\left(\langle X\rangle +\frac{i\langle G\rangle }{2(\Delta P)^2}\langle
P\rangle
\right)\psi_{ss}.
\end{equation}
Of the  four parameters $\langle X\rangle , \langle P\rangle ,
\langle P^2\rangle $, and $\langle G\rangle = \Delta X \Delta P$, only three
are
independent because the uncertainty relation is satisfied.
Therefore,
\begin{equation}
\left(X + iB P\right)\psi_{ss} = C \psi_{ss}  ,~~~
B = \frac{\Delta X}{\Delta P},  ~~~
 C = \langle X\rangle + i B \langle P\rangle .
\end{equation}
Here $B$ is real and $C$ is complex.  These states, $\psi_{ss}(B,C)$, are the
minimum-uncertainty or squeezed states for general potentials
\cite{n1,ss2}.    $B$ can be adjusted to $B_0$ so that the ground
eigenstate of the potential is a member of the set.  Then
$\psi_{ss}(B=B_0,C)=\psi_{cs}(B_0,C)$, are the  coherent states for
general  potentials.  (Recently this method was  connected to a
ladder-operator method for obtaining generalized squeezed states \cite{n2}.)

The natural quantum operators follow the quantum equations of motion.
Therefore,
the
coherent and squeezed state expectation  values of these operators follow the
classical equations of motion.  Intuitively, one can easily understand that if
a
harmonic
oscillator potential is slightly deformed to another potential,  the
appropriate
coherent and squeezed states will still retain their basic properties of
classical
motion, revivals, etc.   What happens here is the opposite.  A general
potential is
classically transformed to a harmonic oscillator-like potential by the
appropriate change
of variables.  This allows the quantum-mechanical coherent and squeezed states
to be
obtained for this pseudo-harmonic oscillator in terms of the physical
operators.

When applied to the Coulomb problem, the natural quantum operators are
\cite{n3}
\begin{equation}
X = \left( \frac{1}{\rho} - \frac{1}{2l(l+1)} \right),
\end{equation}
\begin{equation}
P = p_r = \frac{1}{i}\left( \frac{1}{\rho} + \frac{d}{d\rho} \right),
\end{equation}
where $\rho$ is the dimensionless radius and $p_r$ is the dimensionless radial
momentum.  These operators obey the quantum analogues to the classical
equations of
motion:
\begin{equation}
\dot{X} = \frac{1}{i}[X,H] = -\frac{1}{2}\left\{ \frac{1}{\rho^2},P \right\}
\end{equation}
\begin{equation}
\dot{P} = \frac{1}{i}[P,H] = \frac{l(l+1)}{\rho^2}X.
\end{equation}
The squeezed states obtained from these operators are
 \begin{equation}
\psi_{ss} = [2B\langle 1/\rho \rangle ]^{B+1/2} \Gamma(2B+1)^{-1/2}
       \rho^{B-1} exp[-C \rho],
\end{equation}
\begin{equation}
B = \frac{\langle 1/\rho^2 \rangle}{2[\Delta (1/\rho)]^2},
{}~~~C = B\langle 1/\rho \rangle - i \langle p_r \rangle \equiv u + iv ,
\end{equation}
 When $B= l + 1  $, these states are the coherent states.  The
Coulomb eigenstates are
\begin{equation}
\psi_{nl} = \left[ \frac{\Gamma (n-l)}{2n^4 \Gamma (n+l+1)}\right]^{1/2}
       \exp[-\rho_n/2]\rho_n^l L_{n-l-1}^{(2l+1)}(\rho_n),
\end{equation}
where $\rho_n = \rho/n $ and the L are the associated Laguerre polynomials.
Therefore, the Coulomb ground state ($n=l+1)$ is a coherent state.  (The
decomposition of the coherent and squeezed states into number states is
straight
forward since   $\int_{0}^{\infty}dx~  x^n \exp[-bx] L_k^{(\alpha)}(x)$ is a
standard integral.)
Numerical studies of the coherent
\cite{n4} and squeezed \cite{kos1} states have both been done, and they show
the
appropriate
characteristics, including revivals.  For example, Fig. 1, taken from Ref.
\cite{n4}, shows the early time-evolution of a minimum-uncertainty Coulomb
coherent
state.

The Rydberg wave packets which have been discussed in the literature are
usually
taken as
phenomenological Gaussians.  For large-$n$, these are a good approximation to
the
appropriate squeezed states because, as has been noted
\cite{ryd1,ryd2,ryd3,ryd4},  the
eigenenergies are approximately equally spaced.  In particular, one can compare
the
squeezed Gaussian ($x$ becoming the radial distance $\rho$) and the Coulomb
squeezed state above.  With $p_0=0$ and $C$ real, one can match the highest
point of the
wave packets if $x_0 = (B-1)/C$.  Then the ratio of the zeroth-order to the
second-order
Taylor series about $x_0$ can be matched if $s^2 = (B-1)/C^2$.

Of course, there have
been  discussions of other ``coherent states" for the Coulomb problem
\cite{other}.
An advantage of our point of view is that the states are obtained from a
general method
applicable to arbitrary potentials, and the connection to the associated
squeezed
states is also explicit and general.  In particular, it should be noted that
similar comments about squeezing can be made about other ``coherent" quantum
systems, such as molecules excited by short laser pulses \cite{mole}.

 A final comment on what can be meant by ``coherent" and ``squeezed"
states. For the harmonic oscillator, the concepts ``coherent" and ``squeezed"
are
well-defined and understood by all. But supposing one took a coherent state,
and
distorted it's wave-packet probability form  by $1~\%$ with wiggles at the
edges of the
packet.  Would that packet still ``cohere?"  Of course it would, and one would
not
notice much difference.

This demonstrates a distinction which must be kept in mind. One set of
definitions
of
coherent and squeezed states is composed of  precise mathematical definitions,
be they
from the i) minimum-uncertainty, ii) ladder-operator, or iii)
displacement-operator
points of view. They are  based on mathematical and/or physical
criterion.

On the other hand, there are phenomenological definitions based on approximate
criteria.  Intuitively, these criteria are that the center of the wave packet
follows
the classical motion as well as possible, and that the shape of the wave packet
remains the same, or returns to its original shape periodically, as well as
possible. For historical or simplicical reasons, these approximate definitions
are
very often simply taken to yield Gaussians.  These Gaussians may indeed be good
approximations to more precise definitions, as in this large-$n$ Rydberg case.
Further, since they are packets and not spread-out over a large area, they will
``cohere" and/or ``squeeze" to some, perhaps very good, approximation in the
system
under study.

As long as one keeps these separate uses of the words ``cohere," ``coherent,"
and
``squeezed" clear, then the important thing, the physics, will not become
confused.

I wish to thank Wolfgang Schleich and Carlos Stroud, Jr., for helpful comments.
Also note that related conclusions have been obtained by Bluhm and
Kosteleck\'{y}
\cite{kos1}, based on the supersymmetry-inspired quantum defect-theory
\cite{kos2}.

\newpage

\noindent  Email:  $^a$mmn@pion.lanl.gov

\vspace{1.in}

{\bf Figure 1.} ~~~ The time-evolution of a minimum-uncertainty Coulomb
coherent
state.  $l = 150$, where $B = l+1$.  $C$ (comnplex) is adjusted so that the
particle
starts half-way in time between the classical turning points.  The energy,
$\langle
H \rangle$, is $1/5$ the way up to the continuum from the minimum of the
potential.  The number at the top of the frames indicates the periods of
revolution.

\end{document}